\documentclass[10pt,conference]{IEEEtran}
\IEEEoverridecommandlockouts

\usepackage{cite}
\usepackage{amsmath,amssymb,amsfonts}
\usepackage{algorithmic}
\usepackage{graphicx}
\usepackage{textcomp}
\usepackage{xcolor}
\usepackage{booktabs}

\begin{document}

\title{Beyond Legal Spacing: A Residual-Aware Characterization of Entangling-Zone Spacing in Neutral-Atom Compilation}

\author{\IEEEauthorblockN{Xinyi Li}
\IEEEauthorblockA{\textit{Stevens Institute of Technology}\\
Hoboken, USA \\
xli215@stevens.edu}
\and
\IEEEauthorblockN{Yifeng Peng}
\IEEEauthorblockA{\textit{Stevens Institute of Technology}\\
Hoboken, USA \\
ypeng21@stevens.edu}
\and
\IEEEauthorblockN{Ying Wang}
\IEEEauthorblockA{\textit{Stevens Institute of Technology}\\
Hoboken, USA \\
ywang6@stevens.edu}
}

\maketitle

\begin{abstract}
Neutral-atom processors rely on spatially arranged qubit arrays and parallel Rydberg entangling gates for scalable execution. Their compilers enforce geometric spacing rules for simultaneous gates, yet legal separation does not make residual van der Waals coupling disappear. This paper studies that gap between geometric legality and residual noise by treating entangling-zone spacing as a cross-layer reliability--parallelism variable anchored to experimental neutral-atom geometry. We combine fixed-schedule residual replay, surface-code simulation with matched correlated decoding, and fresh recompilation to connect spacing to physical residual exposure, logical reliability, and makespan cost. The results show that near-floor spacing can produce structured correlated exposure that is visible both at the physical layer and, in the tightest case, after quantum error correction (QEC). Modest geometric slack strongly suppresses this residual contribution, but the timing cost of looser spacing is mediated by placement and scheduling rather than by a simple monotonic slowdown. These findings distinguish hardware legality from residual-noise safety and motivate spacing-aware compiler evaluations that report physical geometry, QEC absorption, and scheduling cost together.
\end{abstract}


\section{Introduction}
Neutral-atom processors are emerging as a promising platform for scalable quantum computation, combining large qubit arrays, reconfigurable atom placement, and Rydberg-mediated entangling gates~\cite{tan2022qubit}. Their reconfigurable geometry is also central to compilation: a compiler must map logical qubits to physical atom locations, move atoms through architecture-specific zones, group entangling operations into Rydberg stages, and enforce spatial constraints among simultaneous two-qubit gates.

Existing neutral-atom compilers encode hardware geometry as a basic architectural setting~\cite{wang2024atomique,tan2025compilation,lin2025reuse}. Values such as zone dimensions, interaction pitch, movement constraints, and minimum separation between concurrent gate pairs are specified as fixed architecture parameters or implementation defaults. These constants then determine which entangling operations may be co-scheduled in each Rydberg stage.

Experimental neutral-atom devices specify physical geometry rules, including a finite blockade radius and a minimum separation between simultaneously gated pairs. In particular, the real-device lower bound of ``no closer than \(10\,\mu\mathrm{m}\)'' and \(R_b\!\approx\!4.3\,\mu\mathrm{m}\) provide a concrete anchor for the entangling-zone spacing~\cite{bluvstein2024logical}.

\begin{figure}[t]
\centering
\includegraphics[width=0.8\linewidth]{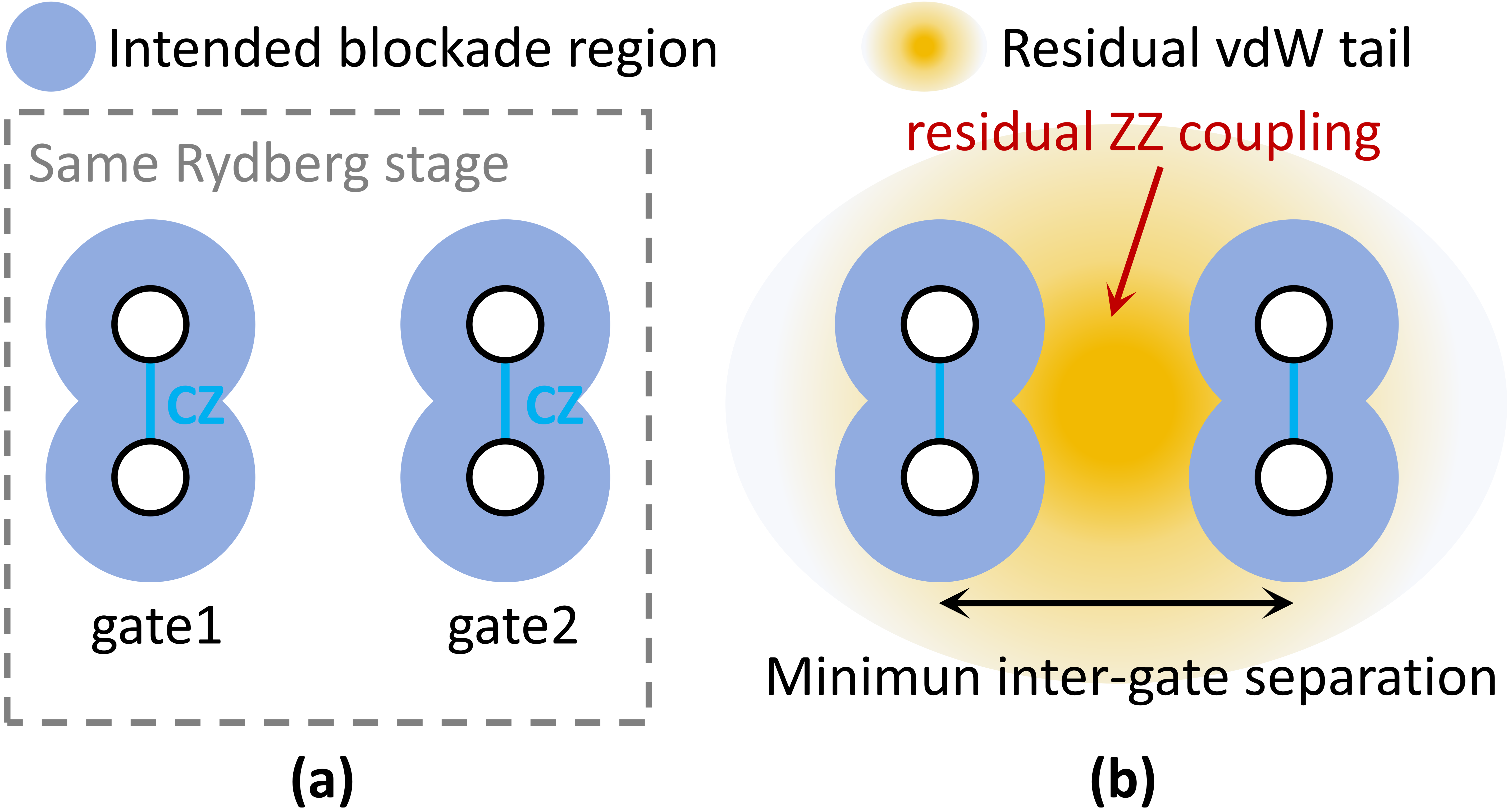}
\caption{Simultaneous neutral-atom entangling gates in a shared Rydberg stage. (a) Intended blockade realizes each CZ within a gate pair. (b) The finite van der Waals tail between distinct pairs produces residual ZZ coupling governed by the minimum inter-gate separation.}
\label{fig:intro-rydberg}
\end{figure}

This lower-bound rule defines gate legality, while residual interactions can persist beyond the intended blockade region. As Fig.~\ref{fig:intro-rydberg} shows, the van der Waals coupling between distinct gate pairs decays continuously as \(1/r^6\). Legal spacings just above the device floor can still induce correlated ZZ-like faults. What remains unclear is how much reliability is gained by moving beyond the lower bound, and how that gain trades off against quantum error correction (QEC) absorption and compiler makespan.

Motivated by this gap, this paper treats entangling-zone spacing as a cross-layer design variable rather than only a geometric legality constraint. We quantify how spacing choices beyond the device floor propagate through three endpoints: physical residual exposure, surface-code logical behavior under a matched correlated decoder, and recompilation cost. Together, these endpoints provide a hardware-anchored characterization of the reliability--parallelism trade-off for neutral-atom entangling zones. Contributions:
\begin{itemize}
  \item Entangling-zone spacing is identified as an overlooked cross-layer issue in which compiler implementations often encode minimum inter-gate separation as a fixed geometry setting, while the device lower bound is only a legality constraint rather than a residual-noise model.
  \item Inter-gate spacing is defined as the swept architectural variable, anchored to a real-device lower bound, and used to evaluate residual ZZ exposure on public neutral-atom schedules with a fixed residual-coupling model.
  \item A surface-code workflow injects residual ZZ faults from simultaneous-gate geometry and decodes the resulting correlated detector-error model.
  \item Fresh recompilation experiments measure how spacing changes both residual exposure and makespan under newly generated schedules for each workload.
  \item The experiments show that spacing beyond the device floor can substantially reduce residual exposure, while QEC absorption and recompilation determine the end-to-end reliability--parallelism trade-off.
\end{itemize}

\section{Background and Setup}\label{sec:setup}
\textbf{Geometry.} Rydberg gates use a finite blockade radius $R_b$, while atoms outside the intended blockade retain a van der Waals tail proportional to $C_6/r^6$. We anchor the geometry to experimental neutral-atom device parameters: $R_b\!\approx\!4.3\,\mu$m and a rule that gated pairs sit no closer than $10\,\mu$m~\cite{bluvstein2024logical}. The two-qubit gate error budget is $0.0046$~\cite{evered2023high}. Table~\ref{tab:compiler-default-spacing} summarizes representative public neutral-atom compiler implementations and their default entangling-zone spacing parameters. Together, they motivate the spacing regimes studied here: the $10\,\mu$m device floor, $11$--$12\,\mu$m near-floor pitches, and $15\,\mu$m-or-larger looser-spacing regimes.

\begin{table}[t]
\centering
\caption{Default spacing parameters exposed by public neutral-atom compiler implementations. Rectangular entries are reported in the implementation's $x{\times}y$ order.}
\label{tab:compiler-default-spacing}
\footnotesize
\begin{tabular}{@{}p{0.25\linewidth}p{0.39\linewidth}p{0.24\linewidth}@{}}
\toprule
Compiler & Target architecture & Default spacing \\
\midrule
Enola~\cite{tan2025compilation} & dynamic field-programmable neutral atoms & $15\,\mu$m \\
ZAC~\cite{lin2025reuse} & zoned neutral atoms & $12{\times}10\,\mu$m \\
PowerMove~\cite{ruan2025powermove} & zoned neutral atoms & $19{\times}15\,\mu$m \\
\bottomrule
\end{tabular}
\end{table}

\textbf{Workloads.} The physical residual audit uses public ZAC schedules~\cite{lin2025reuse} covering $183$ schedule cases and $2{,}597$ multi-gate Rydberg stages. The logical-projection examples use ISING-98, QAOA-20, and a ZAC-compiled surface-code $d=5$ memory instance. The QEC endpoint uses rotated surface-code memory circuits at $d\!\in\!\{5,7\}$, and the recompilation study uses six ZAC benchmarks: QFT-18, GHZ-23, CAT-22, MUL-13, ISING-42, and WSTATE-27.

\textbf{Metrics and method.} \emph{Residual exposure.} We evaluate residual ZZ coupling post hoc on fixed simultaneous Rydberg stages, so the metric isolates residual coupling from rescheduling effects. For two atoms $u$ and $v$ belonging to distinct gated pairs in the same stage, the unwanted van der Waals tail accumulates the residual phase $\phi_{uv}=C_6 t_{\mathrm{gate}}/r_{uv}^6$, where $r_{uv}$ is their separation during the gate and $t_{\mathrm{gate}}$ is the entangling-gate duration. We Pauli-twirl this small coherent over-rotation into a correlated residual-ZZ fault with probability $p_{uv}^{ZZ}=\sin^2(\phi_{uv}/2)$. For a stage $S$, let $\mathcal{P}(S)$ denote all cross-gate atom pairs, namely atom pairs drawn from two different simultaneous intended gates. We report the expected residual exposure
\[
  E_S=\sum_{(u,v)\in\mathcal{P}(S)} p_{uv}^{ZZ}
\]
and the corresponding any-residual probability
\[
  A_S=1-\prod_{(u,v)\in\mathcal{P}(S)}(1-p_{uv}^{ZZ}).
\]
Stage-level residuals are normalized by the two-qubit gate-error budget $p_{2q}=0.0046$ unless otherwise stated.

\emph{Logical error.} Logical-error evaluation uses rotated surface-code memory circuits ($d\!\in\!\{5,7\}$, $X$ basis). Residual ZZ faults are injected between simultaneous entangling pairs in the scheduled Stim circuit. Because these faults can produce detector-error hyperedges, decoding uses BP+OSD directly on the undecomposed correlated detector-error model.

\emph{Makespan.} We recompile each workload at each spacing and report runtime relative to the same workload at $10\,\mu$m.

\begin{figure}[t]
\centering
\includegraphics[width=0.95\linewidth]{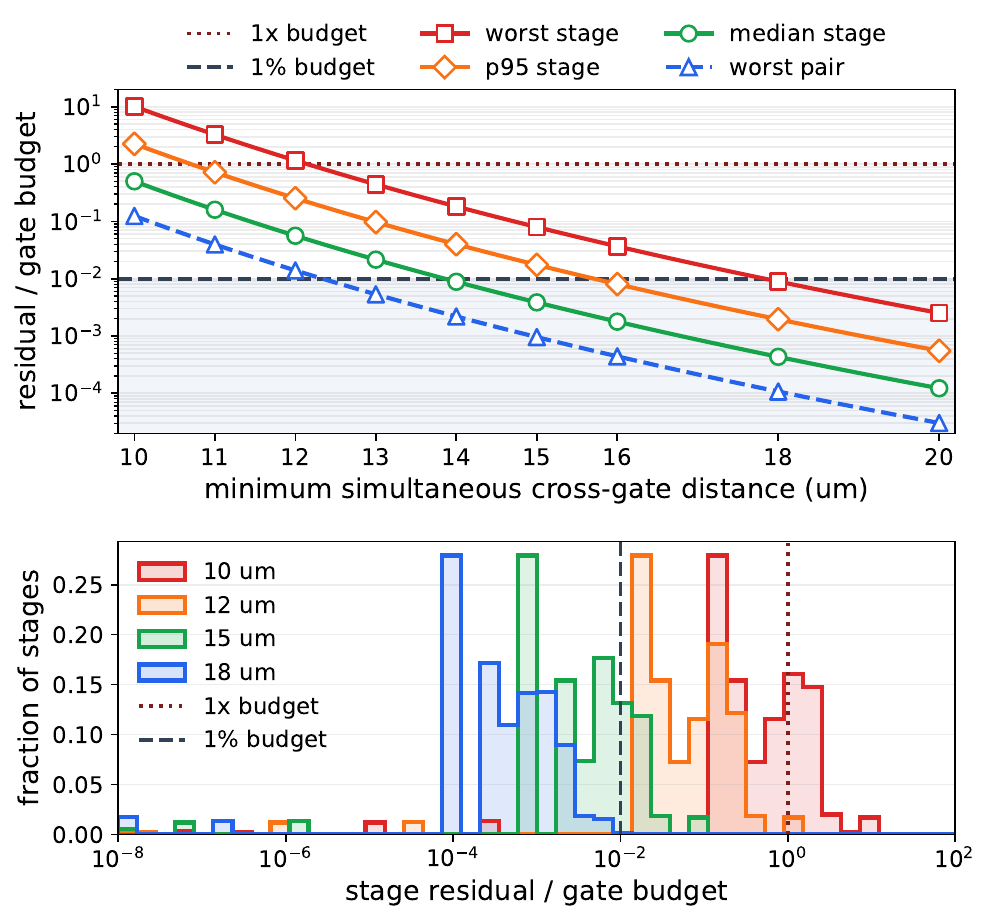}
\caption{Residual exposure in fixed ZAC schedules. Top: stage residual versus inter-gate spacing, normalized to the $0.0046$ two-qubit gate-error budget. Bottom: stage-level residual distributions at selected spacings.}
\label{fig:zac-spacing}
\end{figure}

\section{Residual Exposure vs.\ Spacing}\label{sec:residual}
The upper panel of Fig.~\ref{fig:zac-spacing} shows residual exposure changing sharply near the spacing floor. At $10\,\mu$m, the median stage is $0.50\times$ the two-qubit gate-error budget and the worst stage reaches $10.08\times$. Increasing spacing to $11$--$12\,\mu$m lowers all stage-level curves, but the worst stage remains above budget at $3.26\times$ and $1.15\times$. At $15\,\mu$m, the worst stage falls to $0.080\times$ and the median to $3.86{\times}10^{-3}$. The lower panel shows the same population-level effect: near-floor spacings retain a high-exposure tail, while the $15$ and $18\,\mu$m distributions mostly fall below the $1\%$ budget reference. Thus, legal spacing alone does not make residual exposure negligible. Small geometric slack gives a large reduction, but the sum over many cross-gate pairs can keep stage-level residual exposure visible even when the largest single-pair contribution is small.

\begin{figure}[t]
\centering
\includegraphics[width=\linewidth]{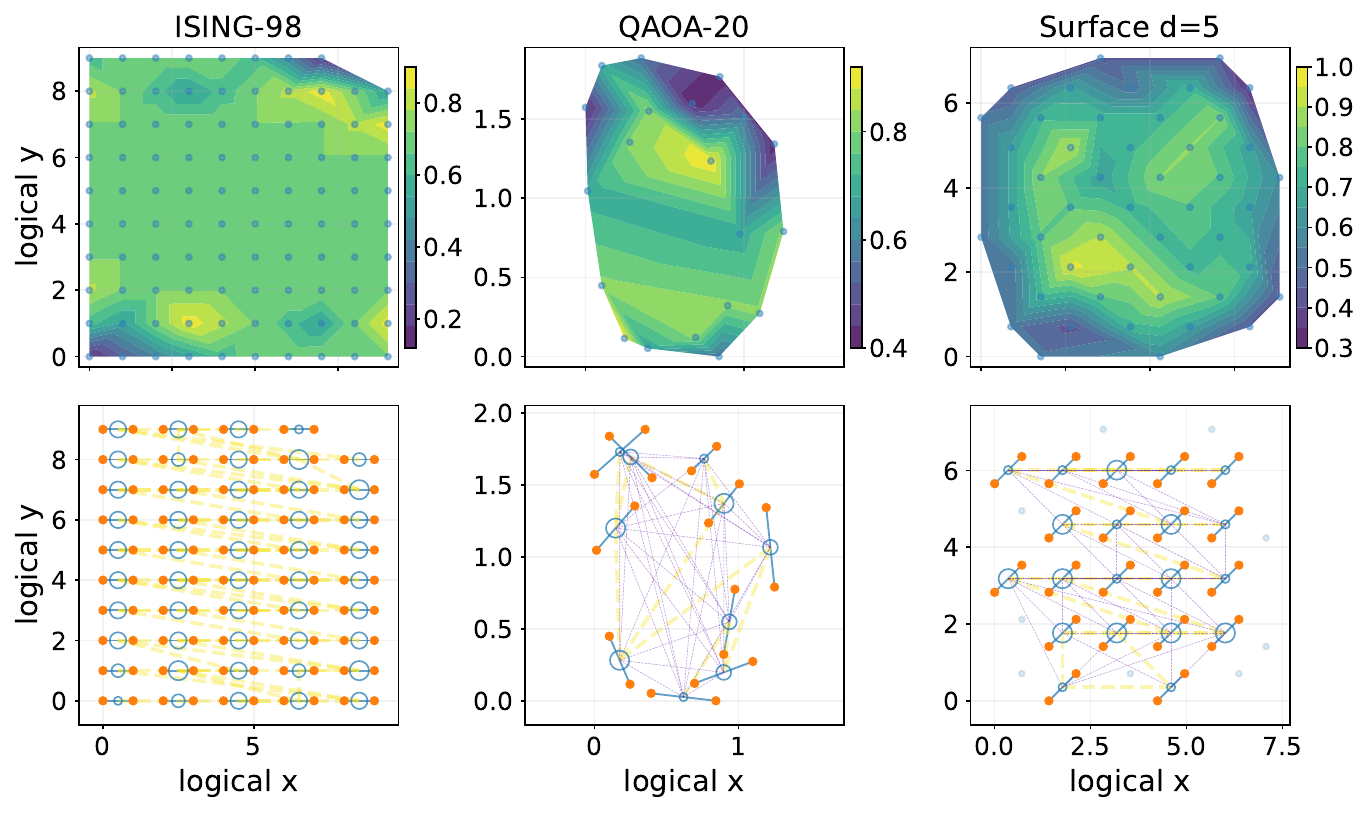}
\caption{Logical-layout projection of physically evaluated residual exposure for ISING-98, QAOA-20, and surface-code $d=5$ memory. Top: accumulated residual exposure over each workload layout. Bottom: the highest-residual compiled stage. Open circles mark active-gate midpoints, with size proportional to gate-level marginal residual exposure. Dashed lines mark the strongest residual-ZZ couplings between simultaneous gates.}
\label{fig:logical-projection}
\end{figure}

Logical-layout projections show that the residual exposure is structured rather than uniform across a workload (Fig.~\ref{fig:logical-projection}). In the upper panel accumulated maps, high-exposure regions appear as localized bands or patches whose shape follows the workload's active entangling structure. The corresponding hotspot stages exceed the two-qubit gate-error budget by $8.91\times$ for ISING-98, $1.25\times$ for QAOA-20, and $3.13\times$ for surface-code $d=5$ memory. The bottom panels explain these peaks: the largest active-gate circles and strongest dashed residual edges occur within the same compiled stage, showing that a small number of densely co-scheduled gate groups can dominate the residual exposure. Thus residual crosstalk should be treated as a structured correlated fault source tied to simultaneous-gate geometry, rather than as an independent error attached separately to each entangling gate.

\section{Logical Error Under QEC}\label{sec:ler}
As quantum noise can degrade application performance~\cite{peng2026quantum,peng2025qsco} and neutral-atom processors advance toward fault-tolerant computation~\cite{bluvstein2024logical}, we investigate whether hardware-specific errors remain consequential after QEC.
Surface-code QEC can suppress many local stochastic faults, but the residual ZZ mechanism considered here induces geometry-dependent correlations between simultaneous entangling gates. Consequently, physical exposure alone is insufficient to determine whether these correlated faults are suppressed by syndrome extraction and decoding or remain a logical-level reliability concern. We therefore inject the residual-ZZ fault model into rotated surface-code memory circuits and decode the resulting correlated detector-error model with BP+OSD.

\begin{figure}[t]
\centering
\includegraphics[width=0.95\linewidth]{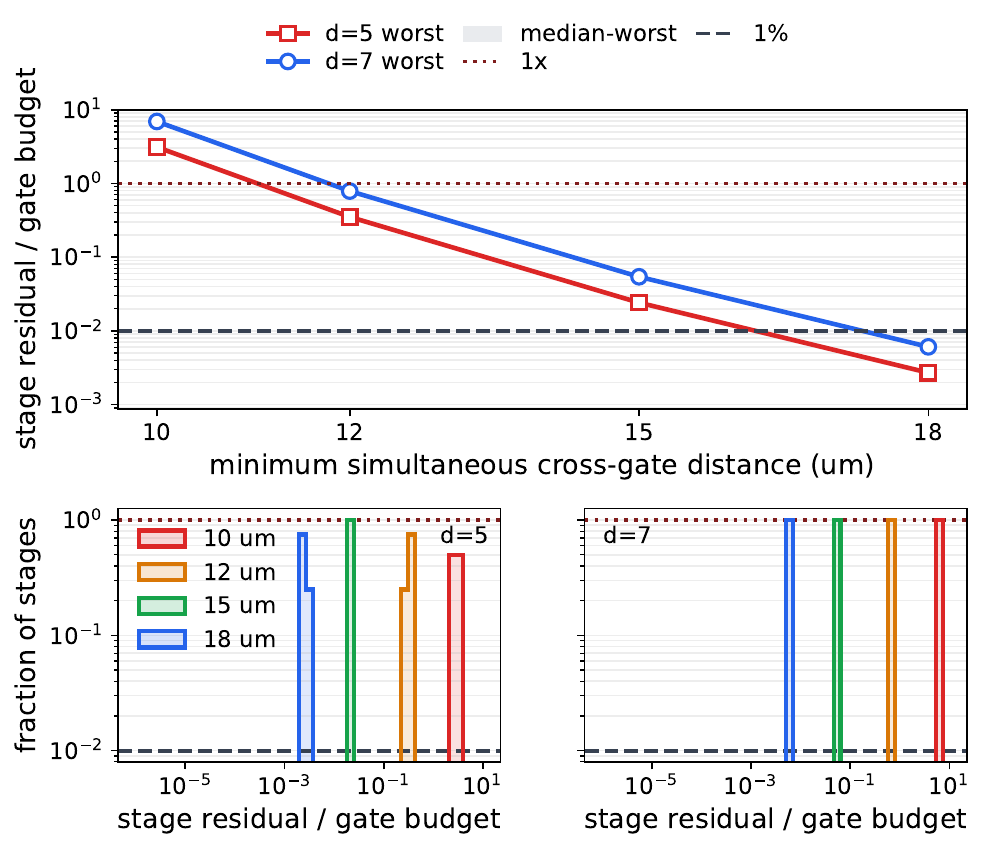}
\caption{Residual exposure in ZAC-compiled surface-code memory. Top: stage residual versus spacing. Bottom: stage-level residual distributions.}
\label{fig:qec-residual}
\end{figure}

Before decoding, the injected residual load in the surface-code memory circuits retains the same spacing dependence observed above (Fig.~\ref{fig:qec-residual}). Exposure is largest at the $10\,\mu$m floor and decreases rapidly as spacing increases. This provides the spacing-dependent correlated fault load used in the logical-error experiment.

\begin{figure}[t]
\centering
\includegraphics[width=0.95\linewidth]{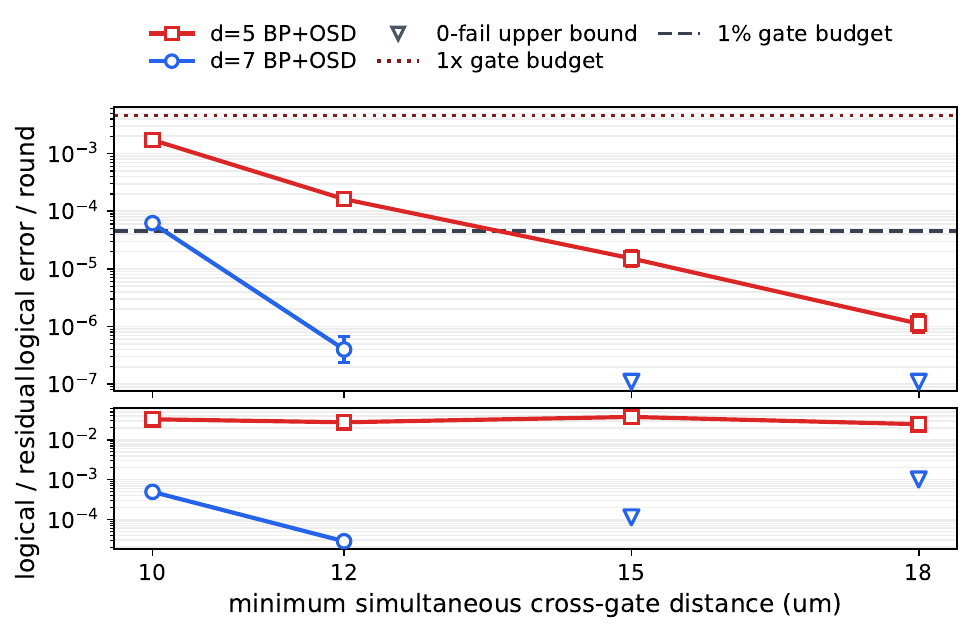}
\caption{Surface-code logical response to residual faults. Top: logical error per round versus inter-gate spacing. Bottom: logical error normalized by the injected residual exposure.}
\label{fig:qec-ler}
\end{figure}

The logical-error results show strong QEC absorption while preserving the spacing trend (Fig.~\ref{fig:qec-ler}). At the $10\,\mu$m floor, residual faults remain visible at the logical level: the error per round is $1.73{\times}10^{-3}$ for $d=5$ and $6.21{\times}10^{-5}$ for $d=7$. For $d=7$, the logical error falls to $4.00{\times}10^{-7}$ at $12\,\mu$m and is not observed in $5{\times}10^6$ shots at $15$--$18\,\mu$m, where we report Wilson upper bounds. The lower normalized panel shows that the larger code converts the same geometry-induced residual mechanism into a much smaller logical fraction. Thus residual crosstalk is a real physical-layer effect that can survive to the logical layer at the device floor, but its impact depends on spacing, code distance, decoder, and fault-correlation structure.

\section{Makespan Cost of Looser Spacing}\label{sec:makespan}
\begin{figure}[t]
\centering
\includegraphics[width=0.95\linewidth]{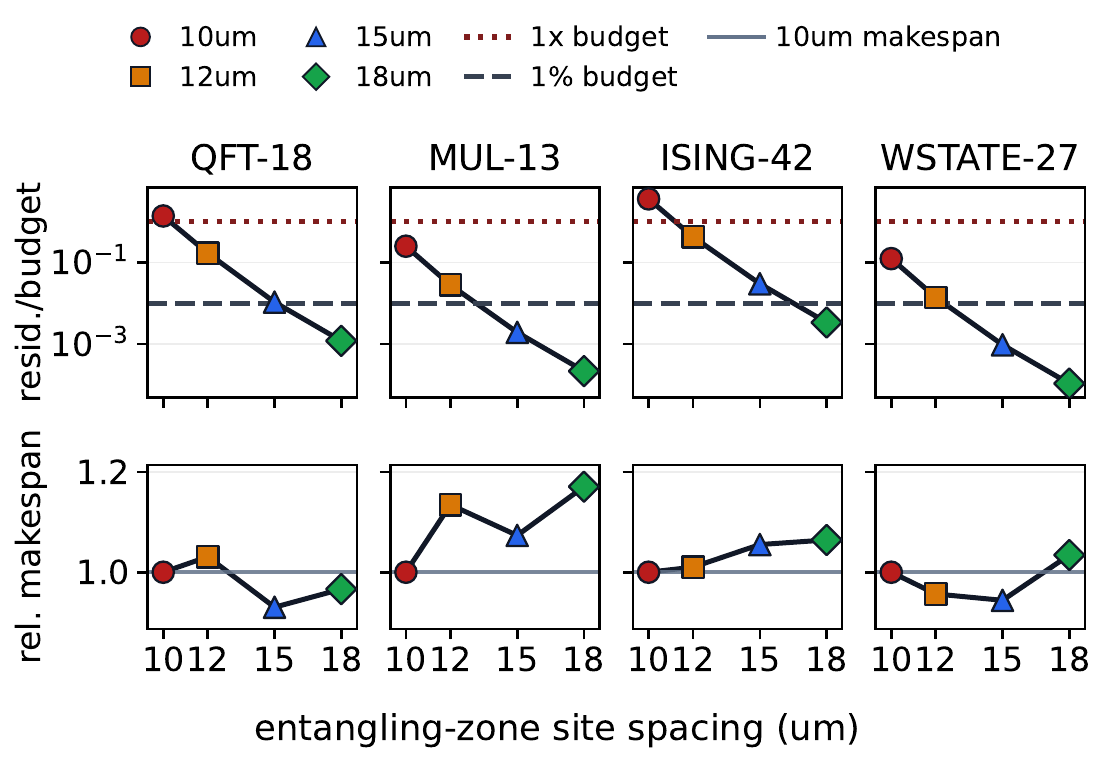}
\caption{Recompiled ZAC workloads at different entangling-zone spacings. Top: worst-stage residual relative to the two-qubit gate-error budget. Bottom: makespan normalized to the circuit's $10\,\mu$m compilation.}
\label{fig:makespan-spacing}
\end{figure}

Recompilation shows an asymmetric trade-off between residual exposure and timing cost (Fig.~\ref{fig:makespan-spacing}). The top panel shows that looser spacing reduces the recomputed worst-stage residual by orders of magnitude, where the median drops from $0.187\times$ the gate budget at $10\,\mu$m to $2.10{\times}10^{-2}$, $1.44{\times}10^{-3}$, and $1.62{\times}10^{-4}$ at $12$, $15$, and $18\,\mu$m. The bottom panel shows a much smaller and workload-dependent makespan response. Median normalized makespan is $1.006\times$ at $12\,\mu$m, $1.004\times$ at $15\,\mu$m, and $1.030\times$ at $18\,\mu$m, while individual circuits range from $0.93\times$ to $1.17\times$. Thus looser spacing reliably reduces residual exposure, but its makespan impact is mediated by the compiler's placement and scheduling choices. Therefore, larger spacing does not induce a simple monotonic slowdown. The reliability--parallelism trade-off is workload- and compiler-dependent.

\section{Discussion and Future Work}\label{sec:discussion}
\textbf{Spacing regimes.} The design implication is that entangling-zone spacing should be exposed and evaluated as an architecture parameter, not hidden as a fixed compiler constant. The device floor is useful for defining legality, but it should not be treated as a residual-noise threshold. Conversely, increasing spacing is not automatically preferable: beyond the near-floor transition region, residual suppression must be weighed against makespan, zone area, and movement overhead.

\textbf{Compiler configurations.} Default entangling-zone spacing is a physical architecture assumption that shapes simultaneous-gate geometry and residual exposure. Two schedules with similar gate count or depth can have different residual exposure if their entangling-zone pitch differs, and compiler comparisons should state this geometry explicitly. We therefore report spacing, simultaneous-gate geometry, and residual-evaluator settings as part of the compiler configuration.

\textbf{QEC-aware scheduling.} The surface-code experiment shows that physical residual exposure is not identical to logical harm: many residual faults are absorbed, while tight spacing can still remain visible at the logical layer. This suggests that future neutral-atom schedulers should evaluate residual events through their detector-level impact, rather than treating all residual pairs as equally important.

\textbf{Future work.} Future work should extend this spacing-centered characterization in three directions: i) calibrate residual-coupling models against device-specific geometry, gate duration, and measured crosstalk across spacing regimes, ii) evaluate spacing choices across larger compiler workloads and QEC layouts where simultaneous-gate geometry changes detector-level correlations, and iii) develop spacing-aware placement and scheduling objectives that trade residual exposure against zone area, movement cost, and makespan under real geometry constraints.

\section{Conclusion}
Entangling-zone spacing exposes a hardware-anchored trade-off between residual crosstalk and parallelism. When the residual ZZ coupling decays as $1/r^6$, legal spacing does not make residual coupling vanish. Near-floor spacings can leave measurable physical exposure, and the $10\,\mu$m floor can still be visible after QEC. At looser spacings the residual contribution is strongly suppressed, while recompilation shows that the makespan cost is workload- and compiler-dependent rather than a simple monotonic slowdown. These results support evaluating spacing, QEC absorption, and scheduling cost together when comparing or designing neutral-atom compilers.

\bibliographystyle{IEEEtran}
\bibliography{reference}

\end{document}